\documentclass{iau} 
\usepackage{graphicx,hyperref}

\title[IAUS330.~~The relation between velocity dispersions and chemical abundances] 
{The relation between velocity dispersions and chemical abundances in RAVE giants}

\author[R.~Smiljanic \& R.~S.~de Souza]   
{Rodolfo Smiljanic$^1$
 \and Rafael Silva de Souza$^2$}

\affiliation{$^1$Nicolaus Copernicus Astronomical Center, Polish Academy of Sciences, Bartycka 18,\\ 00-716, Warsaw, Poland \\ email: {\tt rsmiljanic@camk.edu.pl} \\[\affilskip]
$^2$MTA E\"otv\"os University, EIRSA ``Lendulet'' Astrophysics Research Group,\\ Budapest 1117, Hungary \\email: {\tt rafael.2706@gmail.com}}

\pubyear{2017}
\volume{330}  
\jname{Astrometry and Astrophysics in the Gaia Sky}
\editors{A.~Recio-Blanco, P.~de Laverny, A.~Brown \& T.~Prusti, eds.}
\begin{document}

\maketitle

\begin{abstract}

We developed a Bayesian framework to determine in a robust way the relation between velocity dispersions and chemical abundances in a sample of stars. Our modelling takes into account the uncertainties in the chemical and kinematic properties. We make use of RAVE DR5 radial velocities and abundances together with Gaia DR1 proper motions and parallaxes (when possible, otherwise UCAC4 data is used). We found that, in general, the velocity dispersions increase with decreasing [Fe/H] and increasing [Mg/Fe]. A possible decrease in velocity dispersion for stars with high [Mg/Fe] is a property of a negligible fraction of stars and hardly a robust result. At low [Fe/H] and high [Mg/Fe] the sample is incomplete, affected by biases, and likely not representative of the underlying stellar population.
\end{abstract}

\firstsection 
\section{Introduction}

In a sample of giants from RAVE DR4, \cite[Minchev et al. (2014)]{Minchev14} discovered a decrease in the velocity dispersion of old stars (low [Fe/H] and [Mg/Fe] $>$ +0.4). Comparing with a chemo-dynamical model \cite[(from Minchev et al. 2013)]{Minchev13}, this decrease was interpreted as evidence of the migration of old inner disk stars with cool kinematics. The migration was likely triggered by an early merger. In this work, we re-address the problem of analyzing the relation between chemical abundances and velocity dispersions. We introduce a hierarchical Bayesian approach that models the velocity dispersions without the need of arbitrary data binning (Smiljanic \& de Souza 2017, in prep).

\begin{figure}[t]
\begin{center}
\includegraphics[height=2.6in]{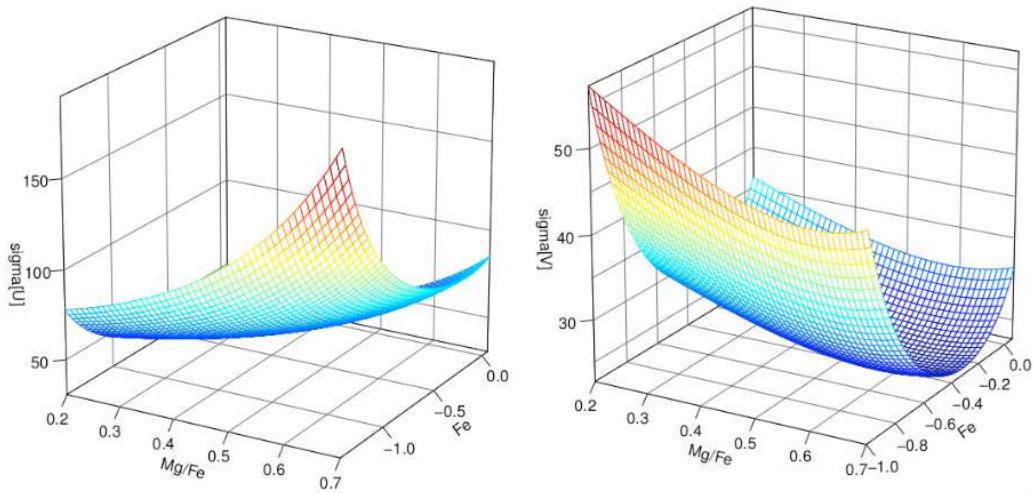}
 \caption{Dispersion of the U (left) and V (right) velocities for giants with [Mg/Fe] $>$ +0.20.}
   \label{fig1}
\end{center}
\end{figure}

\section{Stellar sample}

We use RAVE DR5 \cite[(Kunder et al. 2017)]{Kunder17}. RAVE is a stellar spectroscopic survey that provides radial velocities, atmospheric parameters and abundances for half million stars. We selected RAVE giants ($T_{\rm eff}$ $>$ 4000 K, 1.5 $<$ $\log~g$ $<$ 3.5) with good-quality distances (either based on Gaia parallaxes or from RAVE itself; \cite[Binney et al. 2014]{Binney14}, \cite[Astraatmadja \& Bailer-Jones 2016]{AB16}, \cite[Lindegren et al. 2016]{Lindegren16}), good-quality proper motions, and radial velocities (22242 giants). Here, we present results of the preliminary analysis of a subsample with high-signal-to-noise spectra ($>$ 65) and volume restricted ($|Z| <$ 0.5 kpc, 7 kpc $<$ X $<$ 9 kpc, and $|Y| <$ 1 kpc), i.e., about 4500 stars.

Comparing the atmospheric parameters with isochrones, we noticed that the RAVE giants with [Fe/H] $< -$0.50 tend to be cooler and brighter than expected. This suggests accuracy problems in the analysis of metal-poor stars in RAVE.

\section{Discussion}

We fit the variance of each Galactic velocity component assuming that the logarithm of the variance is a linear function of the velocity. This ensures that the variance is always positive. The MCMC integration is done within R, using JAGS. The uncertainties of the velocities and abundances are fully taken into account in the modelling. 

Our preliminary analysis suggests that, for low-[Fe/H] high-[Mg/Fe] stars (Fig.\ \ref{fig1}), the velocity dispersions are uncertain. The model does not converge well and the credibility intervals are wide. The small number of stars in this regime is likely not representative of the underlying population and does not constrain well the velocity dispersions. Moreover, the low accuracy of the atmospheric parameters also raises questions about the quality of the abundances (Fe and Mg) for these stars. Therefore, we caution that conclusions based on these stars are uncertain and should be seen with care.

\begin{acknowledgment}
We acknowledge E. E. O. Ishida for useful discussions. R. Smiljanic acknowledges support from NCN (grant 2014/15/B/ST9/03981) and from the Polish Ministry of Science and Higher Education. This work has made use of data from the European Space Agency (ESA) mission {\it Gaia} (\url{https://www.cosmos.esa.int/gaia}), processed by the {\it Gaia} Data Processing and Analysis Consortium (DPAC, \url{https://www.cosmos.esa.int/web/gaia/dpac/consortium}). Funding for the DPAC has been provided by national institutions, in particular the institutions participating in the {\it Gaia} Multilateral Agreement.
\end{acknowledgment}

\end{document}